\documentstyle[prd,aps]{revtex}
\tighten
\begin{document}
\draft
\twocolumn[\hsize\textwidth\columnwidth\hsize\csname 
@twocolumnfalse\endcsname
\title{THE NO-DEFECT CONJECTURE IN COSMIC CRYSTALLOGRAPHY}
\author{Jean-Philippe UZAN$^1$  and Patrick PETER$^2$}
\address{D\'epartement d'Astrophysique Relativiste et de Cosmologie,\\
Observatoire de Paris-Meudon, UPR 176, CNRS, 92195 Meudon (France).\\
Email: $^1$Jean-Philippe.Uzan@obspm.fr, $^2$peter@prunelle.obspm.fr}
\maketitle
\begin{abstract}
The topology of space is usually assumed simply connected, but could
be multi-connected. We review in the latter case the possibility that
topological defects arising at high energy phase transitions might
still be present and find that either they are very unlikely to form
at all, or space is effectively simply connected on scales up to the
horizon size.
\end{abstract}
\vskip2pc]

\section{Introduction}

The occurence of topological defects (TD) during the early universe
phase transitions depends only on the topology of the vacuum manifold
$G/H$ when a large symmetry $G$ is broken down to a smaller one $H$
(including $SO(3)\times U(1)$ if the model is to be valid at low
energies). If $G/H$ is disjoint [$\pi_0 (G/H)\not\sim \{\hbox{Id}\}$]
then domain walls must form, while strings or monopoles appear
respectively in the cases $\pi_1 (G/H)\not\sim \{\hbox{Id}\}$ and
$\pi_2 (G/H)\not\sim \{\hbox{Id}\}$~\cite{kibble,book}, with $\pi_n$
the $n^{\hbox{\sevrm th}}$ homotopy group of the vacuum manifold seen
as a topological space. If $G$ is compact and simply connected, which
is usually assumed to be the case for Grand Unified Theories (GUT) in
order to have only one single coupling constant, then the experimental
fact that electromagnetism [$U(1)$] is unbroken leads to the
prediction that monopoles must have formed~\cite{poly}, and with a
number density much too large to be compatible with the evidence that
the universe still exists at all~\cite{preskill}. This observation,
together with the horizon problem, led to the idea of
inflation~\cite{inflation,kolb}. Besides, cosmic strings and domain
walls have been shown to have the ability to generate density
fluctuations that could lead to large scale structure formation,
leaving an observable imprint in the microwave background in so
doing~\cite{lola}. Hence, the possible existence of TDs is not a mere
speculative idea but is seen on the contrary to have cosmological
consequences that are worth investigating (see Ref.~\cite{book} for a
review). So the question of whether they exist or not demands an
answer.

On the other hand, the standard framework in which cosmology is
studied is that of connected universe, an hypothesis essentially based
on a principle of simplicity: since general relativity, being a
differential and therefore local theory, says nothing about the
global, topological aspect of space, it is natural to consider the
universe as endowed with the simplest possible structure and therefore
it is considered simply connected~\cite{kolb}. However, it has been
argued that since quantum gravity possibly allows changes of
topology~\cite{qc1} and because the probability for creation of a
universe decreases with the volume of the universe~\cite{qc2}, a multi
connected space with less volume than a simply connected one is at
least not less probable and could therefore be actually
realized~\cite{ellis}. These models are for the time being constrained
but not excluded~\cite{luminet}. It is the purpose of this letter to
show that a definite observation of a single TD would provide a very
strong constraint on cosmological theories not based on a simply
connected universe. Conversely, if by some other means one were able
to prove the multi-connectedness of the universe, then our result
states that the formation of stable topological defects should be
seriously reconsidered, the probability that we ever observe one being
considerably reduced. This gives an additional insight into the
mechanism for the symmetry breaking, including the dynamics of the
phase transition itself.

It appears therefore that testing the topology of the universe,
i.e. its properties on large, cosmological scales, provides some
information about the nature of particle physics, and in particular
reveals aspects of the possible phenomenology otherwise inaccessible.

The article is organized as follows: in the next section, we derive
the argument according to which TDs are incompatible with a
multi-connected space. More precisely, we argue that long-lived TDs
have a very low probability of being formed, the simplest possible
configurations involving at least two of them in a definite state.
Then, we specialize the discussion to more physical arguments to
conclude on the actual probability that we ever observe a TD in a
multi connected universe.

\section{Defects in nontrivial topological spaces.}

We shall examine in turn the cosmic string, monopole and domain wall
cases, and for strings and wall, we consider only those lying along
incontractible directions, the other ones being unstable against decay
into elementary particles. We discuss them in Sec.~III.

Let us first consider the topology of space in some details in order
to fix the notation. A multi connected space is conveniently described
by its fundamental polyhedron ${\cal P}$ which is convex with a finite
number of faces $\{ {\cal F} \}$ identified by pairs, together with
the holonomy group $\Gamma$ consisting in the collection of
transformations $\gamma$ which carry a face to its homologous face. An
important point concerning the group $\Gamma$ is that none of its
generator (except the identity) can have any fixed
point~\cite{luminet}. Moreover, we shall consider that the phase
transition is driven by a Higgs field $\Phi$ taking values in
$G/H$. For a domain wall resulting of the breaking of a discrete
symmetry, we shall assume $\Phi$ real and taking values in $\{
-1,1\}$, for a string $\Phi$ is a complex field whose phase $\sigma$
should wind $n$ times around some line and for a monopole we assume
$\Phi$ to be a vector in an internal three dimensional space. With
these notations in mind, we can now turn to the different cases.

\subsection{Cosmic Strings and monopoles.}

We begin with the case of a cosmic string lying along an
incontractible direction os the fundamental polyhedron. We assume in
this case, as in the monopole case to be seen later, that the spatial
section of the universe is orientable. This restriction is necessary
for the proof, and can be justified on the ground that one wants $CPT$
to be conserved while $CP$ to be possibly violated (as is observed to
be the case experimentally)~\cite{CPT}. Besides, it is also necessary
in order to have well-defined spinors in all space as is needed in
particle physics where almost all the particle are
fermions~\cite{fermions}. Note this restricts our analysis to
non-twisted field theories~\cite{twist}.

Let us consider the intersection ${\cal L}$ of the fundamental
polyhedron ${\cal P}$ with an arbitrary 2-surface $\Pi$ such that the
phase of the Higgs field responsible for the potential cosmic string
should wind an integer number of time along the line ${\cal L}$:
\begin{equation} {\cal L} = {\cal P}\cap \Pi = \{ A_1 \cdots A_{2p}
\} \end{equation}
for some integer $p$, the points $A_i$ denoting the intersections of
the 2-surface with the edges of the faces. For the corresponding
string to lie along an incontractible direction, it is necessary that
all the faces crossed by ${\cal L}$ have a pair identified in the same
set. In other words, we have that there exists $\gamma\in\Gamma$ such
that $\gamma (A_1 A_2) = (A_k A_{k+1})$ for some $k$. Now, because the
points $A_1$ is identified with $A_k$, being actually physically the
same point, and $A_2$ with $A_{k+1}$, we must have that the phase
variation $\Delta \psi _{A_1 A_2}$ between $A_1$ and $A_2$ should be
precisely equal to that beween $A_{k+1}$ and $A_k$, namely
\begin{equation} \Delta \psi _{A_1 A_2} = -  \Delta \psi
_{A_k A_{k+1}},\end{equation} and hence the total variation along
${\cal L}$ is exactly zero. There is consequently not one single
defect line crossing the 2-surface $\Pi$, the total winding number
having to vanish. Therefore, the only possibility is that of having an
equal number of strings and ``anti-strings'' such that the total
winding number vanishes. Fig.~1 illustrates our point in the case
where the topology is that of a 3-torus.

The monopole case is in fact exactly similar although nearly trivial
since in that case, it should be clear that the winding on the
polyhedron, by definition of the compactedness of the space, must
vanish. Hence there can be only an even number of monopoles trapped in
the fundamental cell; and there should be in fact an equal number of
monopoles and anti-monopole. We discuss the implication of this fact
in Sec.~III.

\subsection{Domain Walls.}

Even though domain walls are conceptually simpler to represent than
the two previous TDs, the proof of their inexistence in multi
connected spaces is slightly more involved and we now turn to it. Also
as in the previous cases, we do not consider walls entirely contained
within the volum of the fundamental polyhedron, since those can
trivially exist and anyway will decay in less than a Hubble time after
their formation.

$\Phi$ has values on each face of the fundamental polyhedron, according
to which we can classify the various faces into subsets:
\begin{equation} {\cal S}^\pm \equiv \left\{ {\cal F} ; \forall M\in
{\cal F}, \langle \Phi\rangle (M)= \pm 1 \right\} \end{equation}
and
\begin{equation} {\cal S}^0 \equiv {\cal P} - ({\cal S}^+ \cup
{\cal S}^- ). \end{equation}
Then because $\forall\gamma\in\Gamma$, one must impose $\langle\Phi
\rangle (\gamma M) = \langle \Psi\rangle (M)$, it should be clear that
the spaces ${\cal S}^{\pm,0}$ are stable under $\Gamma$ and are
connex.

We now define ${\cal S}^1$ as the subset of ${\cal S}^+$ including
only those faces having a boundary in common with a face in ${\cal
S}^0$. Since ${\cal S}^0$ is stable under $\Gamma$, so is ${\cal
S}^1$, and it is not empty. Let us now define ${\cal S}^{1+}={\cal
S}^+ -{\cal S}^1$. Then ${\cal S}^{1+} \subset {\cal S}^+$ and
$\forall\gamma\in\Gamma$, $\gamma{\cal S}^{1+}={\cal
S}^{1+}$. Similarly we define ${\cal S}^2$ as the subset of ${\cal
S}^{1+}$ which is the set of faces having boundaries with faces in
${\cal S}^1$. Again, ${\cal S}^2$ is stable under the action of
$\Gamma$ and is not empty. By induction, it is then possible to define
a series ${\cal S}^{n+}$ such that
\begin{equation} \left\{\begin{array}{l}
{\cal S} ^{n+}\subset {\cal S}^{(n-1)+}\\
\hbox{Card}\,{\cal S}^{n+}\not= 0
\end{array}\right.\end{equation}
since Card~${\cal S}^+<\infty$. Thus, either $\Gamma\sim\{\hbox{Id}\}$
and space is simply connected, or $\exists\gamma\in\Gamma$ with a
fixed point which is impossible since $\Gamma$ is a holonomy group.

Let us turn now to the more complicated case of two domain walls in
the volume. If the two walls cross the polyhedron on different faces,
then the previous analysis applies on two disjoint subsets and the
result is the same. This is also true if they cross the same faces of
the polyhedron while not crossing each other on the face. However the
situation is different when the intersections of the walls and the
face have a common point. In this case, let us define the subsets
\begin{equation} {\cal S}^{00} \equiv \left\{ {\cal F} \in {\cal P};
\Pi_1 \cap {\cal F} \not= \emptyset \ \hbox{and}\ \Pi_2\cap
{\cal F} \not= \emptyset \right\} , \end{equation}
\begin{equation} {\cal S}^{0} \equiv \left\{ {\cal F} \in {\cal P};
\Pi_1 \cap {\cal F} \not= \emptyset \ \hbox{or}\ \Pi_2\cap
{\cal F} \not= \emptyset \right\} , \end{equation}
and
\begin{equation} {\cal S}^\pm \equiv \left\{ {\cal F} ; \forall M\in
{\cal F}, \langle \Phi\rangle (M)= \pm 1 \right\} . \end{equation}
Then the previous proof applying for one wall applies straighforwardly
upon the substitution
\begin{equation} \left\{\begin{array}{l}
{\cal S}^0 \longleftrightarrow {\cal S}^{00}\\
{\cal S}^+ \longleftrightarrow {\cal S}^0.
\end{array}\right.\end{equation}

A generalization to an arbitrary number of walls turned out not to be
feasible in a simple way, so we can only conjecture this result to
apply also in the case of arbitrary many walls. Fig.~2 gives an
intuitive understanding in the case of one domain wall in a 3-torus.

\section{Physical Considerations.}

The basic assumption regarding the possibility of a nontrivial
topology for the universe is that it is made up of a collection of
multi connected spatial sections, all of them obeying Einstein's
equations as well as the cosmological principle of homogeneity and
isotropy. The observable universe itself ${\cal U}$ is identified with
the universal covering of the spatial section ${\cal S}$, so that to
each point in ${\cal S}$ corresponds infinitely many points in ${\cal
U}$ for a multi connected ${\cal S}$. This property is the one that
has been used intensively~\cite{luminet,test} to try and detect
multi-connectedness.

The spatial section has a characteristic length scale $L$ that scales
like the metric. In the case where topology comes from the quantum to
classical gravity transition, one may expect $L$ to be of the order of
the Planck length $\ell_P$ at the Planck time $t_P$ (note however the
possibility of some other length scale, namely the inverse square root
of the cosmological constant, for the actual spatial extension of the
fundamental cell~\cite{cosmocste}; this possibility is not very well
established yet and we shall therefore not consider it any further,
although it should be kept in mind for definite conclusion to be
drawn). Then the leading behavior of $L$ is given by the scale factor
$a(t)$ [normalized to unity at the Planck time]:
\begin{equation} L \sim \ell_P a(t),
\end{equation}
independently of the epoch considered.

For topological defects to be produced at a phase transition, one
needs to investigate the values of the field responsible for the
symmetry breaking over distances larger than its correlation length
$\xi$ which is of the order of the inverse temperature $T_{PT}$ at the
phase transition:
\begin{equation} \xi \sim T_{PT}^{-1} \sim \ell_P a(t)
,\end{equation}
so that the ratio scales like
\begin{equation} {\xi \over L} \sim 1,\label{unity}
\end{equation}
which is valid as long as the defect forming phase transition takes
place at a time where temperature scales like $a(t)^{-1}$. This is
true in particular before the inflationnary phase (if any) and/or prior
to reheating.  Hence we expect in this case only a few TDs to be
formed at the phase transition.

Monopoles can only be formed pairwise with vanishing total index,
which means a quite special field configuration. Thus, it should be
clear that the probability to have monopole forming at the phase
transition is in fact much smaller than that of not forming monopoles
at all. In standard cosmology, one considers various correlation
volumes in which all the possible states are physically realized, and
therefore one expects roughly one monopole per correlation volume on
average. In the compact case however, there is only one such
realization and therefore no ergodic principle can apply. The
conclusion is this case can be based on the anthropic
principle~\cite{anthropic}: as monopoles are a cosmological nuisance
in order for the universe to exist as such still now, it can be
deduced with a good confidence level that monopoles were simply not
formed and the highest probability was realized. If ever we observe
one (and at any energy scale), then we will have to conclude that
either the universe is simply connected, or the phase transition
leading to their appearance took place after reheating (in which case
it is necessary that the correlation length $T_{PT}$ be much smaller
than the cell size $L$), or the length scale at the Planck time was
much greater that the Planck length.

Let us turn to extended defects, namely cosmic strings and domain
walls. Still in the hypothesis of $L=\ell_P$ at $t=t_P$, then
inflation is necessary in multi connected universe models, for
otherwise the size of the universe now would be, assuming scaling of
$L$, roughly 350~km, in obvious contradiction with the
constraint~\cite{luminet} $L\agt 350$~Mpc. But extended defects can be
of two different kinds: either they are contractible in the spatial
topology, in which case they will decay very rapidely and thus have
very little cosmological relevance, or they are aligned along
incontractible directions, so they are topologically preserved as the
universe expands. The latter, as we have seen, are quite improbable
for topological reasons (and we even conjecture walls to be actually
excluded).

Let us see however the implication of an unambiguous observation of a
cosmic string. If inflation was such that the cell size now is much
greater than the horizon, and if the condition of Eq.~(\ref{unity})
was fulfilled, then we expect at most a few strings in the entire
universe and the probability of observing one almost vanishes. Thus,
either we are incredibly lucky, or one of our hypothesis is wrong.
For instance, it could be that $L$ is much greater than the
correlation length at the time of symmetry breaking so that the
ergodic hypothesis applied. However, even in that case, one can
convince oneself that it is not only the field distribution on scales
of the order of the correlation length that matters because the total
configuration must be such that the total winding number vanishes. For
a large number of strings $N$, this requires a phase distribution
whose probability goes like some inverse power of $N$ (depending on
the topology of the spatial section). Hence, we are led back to the
conclusion that space is simply connected.

The last possibility is then that the characteristic lenght $L$ now
exceeds the horizon so that TDs smaller than $L$ did not yet all
decay.  Then multiconnectedness is an irrelevant hypothesis.

\section{Conclusions.}

We have proved that a single topological defect, i.e. a domain wall,
cosmic string or a monopole, cannot appear at a phase transition in a
multi-connected universe. This is based on purely topological
considerations and is therefore completely model-independent. The only
possibility left for TDs to appear are pairwise for cosmic strings and
monopoles, and we conjecture walls cannot form at all. We believe this
is linked with the fact that walls appear as a result of the breaking
of a finite symmetry group, just like the homology group $\Gamma$,
whereas monopoles and strings are created when a continuous group
is broken.

Extended defects can exist however if they are completly inside the
volume of the spatial section. Because this spatial section is small
at the phase transition, we expect only a few such TDs to be formed
anyway, and therefore, if ever created, they decayed almost
instantaneously so that they can't be of any cosmological
relevance. An observation of a TD would thus be a strong indication
that the universe is simply connected.

Another point worth mentionning is the question of monopoles: we have
seen that they are not expected to form, independently of the model
that gives them birth. However, the vaccum manifold in GUT supports
their creation, which is in fact a problem for GUT models, at least
all these models having the standard (and observed) $SU(3)\times U(1)$
as a low energy invariance limit. Inflation is usually invoked to get
rid of these monopole. We see here that a multi-connected spatial
topology does the same work.

Yet, we have implicitely supposed that the spatial characteristic
length $L$ is of the order of the Planck length at the Planck time, an
hypothesis justified by quantum gravity. However, it could be that
this is not a correct hypothesis and that the spatial section can be
anything, in particular the inverse square root of the cosmological
constant has been suggested~\cite{cosmocste}, as long as it scales
with expansion. In this case, our result states that we still can
observe TD if $L$ is at least greater than the horizon size. But in
this case, this means that the universe is simply connected up to the
scale of the horizon, and there is very little chance we can test
multi connectedness observationnally.

Finally we should like to mention yet another possibility, namely that
if by some other means one proves the universe to be multi
connected. That would mean that TD are very unlikely to
exist. However, it would not imply any constraint on particle physics
models. A final point to mention in this case is that such a multi
connected universe would not contain any monopole. This is a way out
for the monopole problem not constraining GUT models.

\section*{Acknowledgments}

We wish to thank B.~Carter, N.~Deruelle, D.~Harari, D.~Langlois and
J.~P.~Luminet for their help in clarifying the subject during various
illuminating discussion.

\end{document}